*Title:* Comparing process-based and constraint-based approaches for modeling macroecological patterns

*Author affiliations:* Xiao Xiao[1, 2, 3, 4], James P. O'Dwyer[5], and Ethan P. White[1, 2, 6, 7]

[1]Department of Biology, Utah State University, Logan, UT 84322-5305

[2]Ecology Center, Utah State University, Logan, UT 84322-5205

[3]School of Biology and Ecology, University of Maine, Orono, ME 04469

[4]Senator George J. Mitchell Center for Sustainability Solutions, University of Maine, Orono, ME 04469

[5]Department of Plant Biology, University of Illinois, Urbana, IL 61801

[6]Department of Wildlife Ecology and Conservation, University of Florida, Gainesville, FL 32611

[7]Informatics Institute, University of Florida, Gainesville, FL 32611

*Corresponding author:* Xiao Xiao, School of Biology and Ecology and Senator George J. Mitchell Center for Sustainability Solutions, University of Maine, Orono, ME 04469. Phone: 435-213-6322. Email address: xiao@weecology.org.

*Authors' email addresses:* xiao@weecology.org, jodwyer@illinois.edu, ethan@weecology.org

## Abstract

Ecological patterns arise from the interplay of many different processes, and yet the emergence of consistent phenomena across a diverse range of ecological systems suggests that many patterns may in part be determined by statistical or numerical constraints. Differentiating the extent to which patterns in a given system are determined statistically, and where it requires explicit ecological processes, has been difficult. We tackled this challenge by directly comparing models from a constraint-based theory, the Maximum Entropy Theory of Ecology (METE) and models from a process-based theory, the size-structured neutral theory (SSNT). Models from both theories were capable of characterizing the distribution of individuals among species and the distribution of body size among individuals across 76 forest communities. However, the SSNT models consistently yielded higher overall likelihood, as well as more realistic characterizations of the relationship between species abundance and average body size of conspecific individuals. This suggests that the details of the biological processes contain additional information for understanding community structure that are not fully captured by the METE constraints in these systems. Our approach provides a first step towards differentiating between process- and constraint-based models of ecological systems and a general methodology for comparing ecological models that make predictions for multiple patterns.

## Introduction

Patterns of biodiversity that are aggregated across large numbers of individuals often take similar shapes across ecosystems and taxonomic groups (Brown 1995). Understanding why such patterns seem to be universal, for example the skewed distribution of individuals among species (the species abundance distribution) (Fisher et al. 1943, McGill et al. 2007) and the uneven allocation of body size among individuals (the individual size distribution) (Enquist and Niklas 2001, Muller-Landau et al. 2006b), is one of the central pursuits of macroecology (Brown 1999, McGill and Nekola 2010).

This task is not trivial because common patterns are often associated with multiple models that have different assumptions about mechanisms yet make similar or even identical predictions (Frank 2014). For example, more than 20 models exist for the species-abundance distribution (SAD) all making realistic predictions with many rare species and a few abundant ones (often known as a hollow-curve), but with mechanisms ranging from purely statistical to population dynamics to resource partitioning (Marquet et al. 2003, McGill et al. 2007). Moreover, many macroecological patterns are not independent. For example, the species-area relationship at small spatial scales can be derived from the shape of the SAD and the level of intraspecific aggregation (Harte 2011, McGill 2011), while the SAD itself can be obtained as a spatially autocorrelated sample from the regional species pool (McGill 2011). This combination of equivalent models with different processes and interrelated patterns makes determining process using a single pattern challenging and instead calls for unified theoretical frameworks that are capable of capturing multiple patterns as well as their intercorrelations with a minimal set of assumptions (Marquet et al. 2014).

Theories that have been proposed for macroecological patterns tend to fall into two conceptually distinct categories (Brown 1999, Frank 2014). Similar patterns may arise directly from fundamental ecological processes if the same processes dominate across multiple systems. Theories in this category include the theory of island biogeography (MacArthur and Wilson 1967), which explains the species richness on islands as the equilibrium between immigration and extinction, and the neutral theory of biodiversity (Hubbell 2001), which shows that demographic stochasticity can lead to community-level diversity patterns. Alternatively, patterns may arise as emergent statistical phenomena with forms determined primarily by some set of numerical constraints on the system (Frank 2014), where processes operate only indirectly through their effects on the constraints. Theories built on constraints include the feasible set (Locey and White 2013), and recent applications of the Maximum Entropy Principle to ecology (Shipley et al. 2006, Dewar and Porté 2008, Harte 2011). Neither of these approaches relies on the operation of specific processes but instead on the fact that many possible combinations of processes and states of the system produce similar empirical outcomes (Frank 2014).

In this study we examined two theoretical frameworks, the Maximum Entropy Theory of Ecology (METE) (Harte 2011) and the size-structured neutral theory (SSNT) (O'Dwyer et al. 2009), which are two of the most comprehensive theories in macroecology. Both theories are able to predict two distinct sets of patterns, those of biodiversity as well as body size and energy use. METE is a constraint-based theory, where patterns arise as the most likely (least biased) state of a community constrained by a set of state variables, such as species richness, the total number of individuals, and the total energy consumption across all individuals. SSNT is an extension of the neutral theory of ecology (Hubbell 2001) and is a process-based theory, where the patterns arise as the steady state of a dynamic system governed by individual birth, death, and growth in size. Both theories make predictions for multiple patterns of biodiversity as well as biomass and energy use, providing a multifaceted characterization of community structure.

We evaluated two existing models of METE and two models that we derived for SSNT, to explore whether community structure in biodiversity and body size can be adequately captured by constraints or processes. One of the METE models, ASNE (Harte 2011; see Methods for details), has been shown in previous studies to have mixed performance among its predictions (Newman et al. 2014, Xiao et al. 2015), while the other models have not been thoroughly tested with empirical data. Using data from 76 forest communities we examined the models' ability to characterize three major macroecological patterns, and compared their performance using a single joint distribution that encapsulates these and other predictions as marginal or conditional distributions. Direct comparison of multiple models from the two theoretical frameworks, using a large number of datasets and multiple empirical patterns, allows strong inference to be made about the relative performance of the models and, by extension, the ability of current constraint-based and process-based approaches to characterize community-level macroecological patterns of diversity and body size.

## Methods

### Theoretical frameworks

This section briefly outlines the underlying assumptions of METE and SSNT, and the specification of the two models under each theory. For mathematical configurations and derivation of the predicted patterns, see **Appendix A**.

#### *Maximum Entropy Theory of Ecology (METE)*

The Maximum Entropy Theory of Ecology (METE) is a theory built on the maximum entropy principle (MaxEnt; Jaynes 2003). MaxEnt states that the least biased state of a system is the one with the highest information entropy (Shannon entropy). Given a set of constraints that the system has to satisfy, this state can be obtained by optimization using the method of Lagrange multipliers, with no tunable parameters besides the constraints.

Among existing applications of MaxEnt to ecology (e.g., Shipley et al. 2006, Pueyo et al. 2007, Dewar and Porté 2008), METE is arguably the most comprehensive, encompassing three distinct branches of ecological patterns – the spatial distributions of individuals and species, the distributions of individuals among species and higher taxonomic ranks, and the allocation of body size and energy use at different taxonomic levels. We examined two existing models of METE, ASNE (Harte et al. 2008, Harte 2011), where the acronym stands for Area, Species, Number of individuals, and Energy, and the newly developed AGSNE (Harte et al. 2015), where the additional "G" stands for Genera or other higher taxonomic ranks (family, order, etc.). In this study we focused on the non-spatial patterns in ASNE and AGSNE, which are predicted independently from the spatial patterns. In non-spatial ASNE, the allocations of individuals and of body size within a community are regulated by three state variables: species richness $S$, total abundance $N$, and total metabolic rate within the community $E_{METE}$. Non-spatial AGSNE requires an additional input $G$ for a higher taxonomic group, which we took to be the number of genera within the community.

*Size-structured neutral theory (SSNT)*

Size-structured neutral theory (SSNT) is an extension of Hubbell's neutral theory of ecology (NTE; Hubbell 2001). In NTE, macroecological patterns emerge as the steady state of the community where individuals go through the processes of birth, death, and speciation. SSNT introduces a size component into NTE, where the size of each individual increases through time. Ontogenetic growth thus introduces variation in individual sizes, and also variation in the average size and total biomass across different species. The structure of the community in SSNT is governed by the forms and values of the three demographic parameters $b$ (birth rate), $m$ (mortality rate), and $g$ (rate of growth).

We examined two realized models of SSNT. In the simplest model (SSNT_N, with "N" for neutral), all three demographic parameters are assumed to be constant for all individuals regardless of their species identities or other individual characteristics. This is called the "completely neutral case" in (O'Dwyer et al. 2009). Note that while the assumption of $b$ and $m$ being constant holds regardless of the unit used for size, $g$ can only be constant in one particular set of size units, e.g., constant growth in diameter as a function of current diameter does not translate into constant growth in cross-sectional area or volume as a function of current area or volume. In SSNT_N, we made the intentionally naïve assumption that $g$ was constant across individuals when measured as the increase in diameter $D$ (i.e., $g(D) = \frac{dD}{dt} = constant$).

In the second model, termed SSNT_M where M stands for metabolism or metabolic theory, we incorporated insights from the metabolic theory of ecology (MTE; Brown et al. 2004), and made the more realistic assumption that $g$ was a function of size, while $b$ and $m$ were still held constant. Specifically, MTE predicts that a plant's growth rate measured as increase in biomass is proportional to the plant's metabolic rate (Enquist et al. 1999, West et al. 1999, Muller-Landau et al. 2006a), which translates into constant growth rate when size is measured in units of diameter raised to the power of 2/3, $D^{2/3}$.

*Macroecological patterns*

All four models can predict the same set of three major macroecological patterns: the species-abundance distribution (SAD; distribution of individuals among species), the individual size distribution (ISD; distribution of body size among individuals regardless of their species identity), and the size-density relationship (SDR; relationship between average body size within each species and the abundance of the species)(Cotgreave 1993). AGSNE is also able to predict higher-order patterns, such as the distribution of individuals and body size among genera, which we do not examine in this study (but see Harte et al. 2015).

Table 1 summarizes the predicted forms of the patterns in the four models. $\lambda$'s in ASNE and AGSNE are Lagrange multipliers (Jaynes 2003), determined by the state variables $S$, $N$ and $E_{METE}$ in ASNE (see Harte 2011 and Xiao *et al.* 2015 for detailed derivation), and by $G$, $S$, $N$ and $E_{METE}$ in AGSNE (Harte et al. 2015; **Appendix A**). $\tau$'s in the SSNT models are ratios of the demographic parameters, and are also fully determined by the variables $S$, $N$, and a measure of total body size $E_{SSNT}$, with $E_{SSNT_N} = \sum D_i$ and $E_{SSNT_M} = \sum D_i^{2/3}$ (see **Appendix A**). Note that patterns of body size (ISD and SDR) predicted by the METE models are of the same unit as metabolic rates ($B$), which scales with size in trees with good approximation as the square of diameter ($D$): $B \propto D^2$ (West et al. 1999), while the basic unit of size in SSNT is $D$. For the purpose of comparison, we converted patterns of size from the models into the same units. The ISDs in ASNE and AGSNE were converted to unit of $D$. The SDRs predicted by these two models do not have simple analytical forms in unit of $D$, so we converted the predictions of SSNT_N and SSNT_M to unit of $B$ (i.e., $D^2$) instead (Table 1; **Appendix A**).

<u>Data</u>

We used forest census data to empirically evaluate the models. This type of data consistently includes individual level size measurements, allowing the compilation of large numbers of communities with the necessary information for fitting and evaluating the models. Forest data sample all individuals of every species down to a certain minimum size, thus avoiding issues with not detecting juvenile organisms (other than those below the minimum size), which may bias the empirical size distributions. In addition, determinately growing organisms (e.g., birds and mammals) often exhibit multimodal ISDs (Ernest 2005, Thibault et al. 2011), whereas the ISDs for trees are in general monotonically decreasing (Enquist and Niklas 2001, Muller-Landau et al. 2006b), and therefore consistent with the qualitative form predicted by the four models (Table 1).

We combined the data compiled by (Xiao et al. 2015), which encompassed 60 forest communities worldwide, with data on 20 additional communities from (Bradford et al. 2014). All communities have been fully surveyed with species identity and measurement of size (diameter or equivalent) for each individual above community-specific size thresholds (ranging from 10mm to 100mm). In cases where multiple surveys are available for a community, we used those from the most recent survey unless otherwise specified. We excluded individuals that were not identified to genus or species, and removed four communities in (Xiao et al. 2015) from our analysis where more than 10% of the individuals were not identified to genus. We also excluded individuals that were dead or missing size measurements, as well as those with sizes below or equal to the specified threshold, since not all individuals in these size classes were included in the surveys. Overall the compilation encompasses 76 communities with 2030 species and 378806 individuals from 4 continents (Asia, Australia, North America, and South America) (Table 2).

Analyses

We applied the four models to each empirical community, and examined their abilities to characterize community structure in abundance and body size. Diameter values in each community were rescaled as $D = D_{\text{original}} / D_{\text{min}}$, where $D_{\text{min}}$ is the diameter of the smallest individual in the community after the exceptional individuals were excluded (see 2. Data), so that $D$ has a minimal value of 1 in each community following METE's assumption (see Harte 2011). While SSNT does not make such an assumption on minimal size, it predicts the ISD to be an exponential distribution (Table 1), the fit of which is unaffected by this transformation of $D$. Multiple branches from the same individual were combined to determine the basal stem diameter with the pipe model, which preserves the total area as well as the metabolic rate of the branches (Ernest et al. 2009). Predictions of the models in each community were obtained with the variables *S*, *N*, and $E_{METE} = \sum_i D_i^2$ for ASNE, *G*, *S*, *N*, and $E_{METE}$ for AGSNE, *S*, *N*, and $E_{SSNT_N} = \sum_i D_i$ for SSNT_N, and *S*, *N*, and $E_{SSNT_M} = \sum_i D_i^{2/3}$ for SSNT_M.

As an overall measure of model performance, we define the joint distribution $P(n, D_1, D_2, \ldots, D_n)$ as the probability that a species randomly selected from the community has abundance *n*, while individuals within the species have diameter $D_i$'s with *i* ranging from 1 to *n*. This distribution combines all three macroecological patterns, where the SAD is the marginal distribution of *n* with $D_i$'s integrated out from $P(n, D_1, D_2, \ldots, D_n)$, the ISD is the marginal distribution of $D_i$, and the SDR is the expectation of the conditional distribution of $D_i$ given *n*. The form of this joint distribution predicted by each of the four models is listed in Table 3 (see **Appendix A** for derivations).

We first compared the performance of the four models using the likelihood of $P(n, D_1, D_2, \ldots, D_n)$ in each community, then examined each of the three macroecological patterns individually. To quantify the predictive power of the models, we converted the SAD and the ISD into rank values, where the abundance of species or the diameter of individuals were ranked from the highest to the lowest, and the value at each rank was compared to the models' predictions. For example, for the SAD we compared the predicted versus observed abundances of the most abundant species in the community, the second most abundant species, all the way down to the least abundant species (Harte 2011, White et al. 2012, Xiao et al. 2015). For the SDR, we compared the observed average metabolic rate ($\acute{D^2}$) within each species to those expected from the models. The explanatory power of the models for each pattern was quantified using the coefficient of determination $R^2$:

$$R^2 = 1 - \frac{\sum_i [\log_{10}(obs_i) - \log_{10}(pred_i)]^2}{\sum_i \left[\log_{10}(obs_i) - \overline{\log_{10}(obs_i)}\right]^2}$$

where $obs_i$ and $pred_i$ were the $i$th value of abundance or size (diameter for the ISD, metabolic rate for the SDR) in the observed and predicted ranked distributions, respectively. Note that it is possible for the coefficient of determination to be negative, which indicates that the prediction is worse than taking the geometric mean of the observed values. Finally, we examined if the empirical patterns were significantly different from the models' predictions by bootstrap analysis (Clauset et al. 2009, Connolly et al. 2009, Xiao et al. 2015), where we generated random samples from the predicted patterns and quantified their deviation from the predictions ($pred_i$'s) using both $R^2$ and the Kolmogorov-Smirnov statistic, which were compared with empirical deviations (**Appendix B**).

Python Code to fully replicate our analyses is deposited in the Dryad Digital Repository (doi:10.5061/dryad.93ct6).

**Results**

The log-likelihoods of the joint distribution $P(n, d_1, d_2, \ldots, d_n)$ of the SSNT models are higher than those of the METE models in all 76 communities (Fig. 1), which implies that SSNT models consistently do a better job characterizing the overall community structure in the allocations of individuals and of body size. Comparing models under the same theoretical framework, the log-likelihood of AGSNE is higher than that of ASNE in all 76 communities, while the log-likelihood of SSNT_M is higher than SSNT_N in 59.

Further examination of individual patterns show that the models predict nearly identical forms for the SAD (i.e., upper-truncated log-series in ASNE, near log-series in AGSNE (Harte et al. 2015), untruncated log-series in SSNT_N and SSNT_M; see Table 1), which not surprisingly translates into equally good performance when evaluated with empirical data (Fig. 2, first column). All four models are also able to characterize the ISD reasonable well with high predictive power ($R^2_{ASNE}$ = 0.89, $R^2_{AGSNE}$ = 0.90, $R^2_{SSNT_N}$ = 0.86, $R^2_{SSNT_M}$ = 0.96). However, three of the models show systematic deviations for the largest individuals, with the two METE models tending to over predict the size of the largest individuals and SSNT_N tending to under predict (Fig. 2, second column).

The discrepancy of the two sets of models lies mainly in their predictions of the correlation between individual body size and species abundance. The two METE models both predict that individual body size depends on the species that the individual belongs to – in ASNE average individual body size is negatively correlated with species abundance (Harte 2011), while in AGSNE it is negatively correlated with both species abundance and number of species within genus (Harte et al. 2015). The SDR predicted by ASNE has been shown to be unrealistic in plant communities (Newman et al. 2014, Xiao et al. 2015), and our results show that AGSNE improves ASNE's prediction only marginally ($R^2_{ASNE}$ = -2.11, $R^2_{AGSNE}$ = -2.00). SSNT, on the other hand, predicts that there is no correlation between individual size and species characteristics, leading to better, but still far from good, agreement with empirical data for the SDR ($R^2_{SSNT_N}$= 0.09, $R^2_{SSNT_M}$ = 0.02). These results are robust when the two models are examined in each of the 76 communities individually (Fig. 3) – the four models yield nearly identical $R^2$ values for the SAD and comparable $R^2$ values for the ISD across communities (with SSNT_M having the highest predictive power on average), while the two SSNT models consistently outperforms the two METE models for the SDR.

The bootstrap analysis (**Appendix B**) shows that the discrepancy between the models' predictions and the observations for the ISD is almost ubiquitously higher than expected from random sampling in all four models. This suggests that none of the models is able to fully capture the observed variation in the size distributions of individuals, despite their high predictive power for this pattern. A similar pattern is observed for the SDR predictions for the METE models. In contrast, despite the low predictive power, SDR predictions for the two SSNT models suggest that the majority of the communities are indistinguishable from random samples of the predicted pattern (Figs. B3, B4). This implies that SSNT's prediction of no correlation between species abundance and individual body size is more or less accurate.

**Discussion**

In this study we compared the performance of the Maximum Entropy Theory of Ecology (METE) (Harte 2011) and the size-structured neutral theory (SSNT) (O'Dwyer et al. 2009), two of the most comprehensive theories to date in macroecology, using two realized models for each. Both theories attempt to unify multiple aspects of community structure under a single theoretical framework, predicting patterns of biodiversity as well as patterns of energy consumption and body size. Using data from 76 forest communities worldwide, we showed that the two models of SSNT consistently provide a better characterization of overall community structure than the two models of METE (Fig. 1). This disparity results primarily from the ability of SSNT_N and SSNT_M to more accurately characterize the relationship between species abundance and body size distributions within species, while the predictions of ASNE and AGSNE on this relationship deviate from empirical patterns (Newman et al. 2014, Xiao et al. 2015).

By comparing multiple competing models on multiple predictions simultaneously using an extensive set of data, our study achieves the strongest level of model evaluation suggested by McGill et al. (2006), and provides insights into the role of the underlying mechanisms of the theories. In METE, the macroecological patterns arise as the most likely state of the system assuming that the system is constrained by a small number of state variables. METE makes no explicit assumptions about ecological processes, leaving their influence to operate indirectly through their potential effects on the values of the state variables. In SSNT, patterns emerge directly from the interactions of the demographic processes including birth, death, and growth. The fact that SSNT performs better than METE suggests that the demographic processes contain meaningful information that helps to characterize the patterns, the effect of which is not currently captured by the state-variable based models.

While the differences between the models are important, the fact that all four models are capable of adequately characterizing the shapes of the SAD and the ISD across a large number of communities with simple assumptions and limited inputs is impressive. Moreover, the maximum likelihood parameters for SSNT_N and SSNT_M are also fully determined by $S$, $N$, and $E$ (see **Appendix A**), so that these variables serve as summary statistics for the demographic parameters. These results imply that METE and SSNT contain overlapping information. While these demographic processes explain a higher proportion of the variation in the empirical patterns, their effects are likely to at least be partially channeled through the constraints.

Our study is one step towards the goal of disentangling the effects of different mechanisms on macroecological patterns. While we have adopted model comparison for stronger inference, we do not advocate rejecting the theoretical framework (METE) with the poorer fitting models or its underlying constraint-based view as a potential explanation for patterns. There are three reasons for being cautious about over interpreting these results. First, METE and SSNT are general theories built on first principles, while our conclusions are limited to their current realized models based on specific assumptions. Models under the same theoretical framework yield different predictions with different assumptions and inputs, which can be evaluated with empirical data and improved with additional information. This is demonstrated in our study by comparing the two models from the same theory – with an additional constraint $G$, AGSNE has consistently higher likelihood than ASNE (Fig. 1), while incorporating information from the metabolic theory in SSNT_M eliminates the systematic deviation in the predicted ISD (Fig. 2). Future models will likely be developed with alternative implementations leading to new and/or improved predictions.

Second, our inference is limited by the scope of the data. Though the models have the potential to be applied to a wide variety of systems, we focused exclusively on trees, where data of full surveys are readily available with species identity and body size for all individuals. While our results are consistent across forest communities of different types and sizes (Fig. 3), it remains to be seen if they can be generalized to other taxa. Previous studies suggest that the size distributions predicted by ASNE and AGSNE may be more accurate when applied to invertebrates (Harte 2011, Harte et al. 2015). Third, patterns that can be unified under the same theoretical framework do not necessarily have to arise from the same underlying mechanism. Indeed, there is increasing evidence that the SAD is driven by statistical properties of the system (White et al. 2012, Locey and White 2013, Blonder et al. 2014), while patterns that show spatial or taxonomical variation, such as the patterns of body size, are more likely to be tied to ecological processes (Blonder et al. 2014).

One weakness prevalent across all four models is their inability to characterize the SDR, the relationship between species abundance and the body size of individuals within species, despite their success in independently predicting the distribution of individuals among species and the allocation of body size among individuals. Our results agree with previous studies showing that the SDR exhibits significant variation at the local scale (Lawton 1990, Cotgreave 1993), not strongly abundance-dependent as the METE models predict. The prediction of the SSNT models that the SDR results from random draws is more in line with empirical observations (**Appendix B**), but they too lack predictive power (Fig. 2). While part of the variation may result from the limitation of data we used, e.g., species having different growth rates were surveyed at different life stages, it could also indicate species-specific size biases in resource use (White et al. 2007). One potential remedy that may improve the predicted SDR as well as lead to additional predictions is to take an approach alternative to the two that we have addressed, and model macroecological patterns by directly stacking models of individual species. This approach has shown promise in predicting other patterns, such as biodiversity across space (Guisan and Rahbek 2011, D'Amen et al. 2015). Similar models could potentially be developed to model the abundance and body size of species based on their traits, and to obtain the macroecological patterns from the species-level predictions. However, such models will likely sacrifice parsimony for accuracy, and require a lot more parameters than the models that we examined.

Another potentially fruitful route to push the two theories forward is to unify the constraint- and the process-based approaches, which have generally been adopted by distinct theories but do not necessarily have to be mutually exclusive. Results of our study imply that part of the effects of the demographic processes propagate through the constraints, while other studies (e.g., Haegeman & Etienne 2010) state that different configurations for the same set of constraints can often be tied to (and may eventually be informed from) process-based mechanistic models. Indeed, an exciting new model is being developed where the maximum entropy principle is combined with demographic processes to characterize not only the steady state but also temporal dynamics of a system (Umemura and Harte 2015). The attempts to model ecological systems completely with constraints or processes may thus represent two extremes of a continuous spectrum, along which multiple models exist that lean towards one approach or the other, yet all provide adequate characterization of the system if properly formulated. We look forward to future studies that combine new theoretical development with strong empirical tests to further elucidate the entangled effects of constraints versus biological processes in structuring ecological systems.

**Acknowledgements**

We thank B. Blonder and J. Harte for their constructive reviews, and members of the Weecology Lab for feedback on this research. R. K. Peet provided data for the North Carolina forest plots. The Serimbu (provided by T. Kohyama) and Shirakami (provided by T. Nakashizuka) datasets were obtained from the PlotNet Forest Database. The eno-2 plot (provided by N. Pitman) and DeWalt Bolivia (provided by S. DeWalt) datasets where obtained from SALVIAS. The BCI forest dynamics research project was made possible by National Science Foundation grants to S. P. Hubbell: DEB-0640386, DEB-0425651, DEB-0346488, DEB-0129874, DEB-00753102, DEB-9909347, DEB-9615226, DEB-9615226, DEB-9405933, DEB-9221033, DEB-9100058, DEB-8906869, DEB-8605042, DEB-8206992, DEB-7922197, support from the Center for Tropical Forest Science, the Smithsonian Tropical Research Institute, the John D. and Catherine T. MacArthur Foundation, the Mellon Foundation, the Small World Institute Fund, and numerous private individuals, and through the hard work of over 100 people from 10 countries over the past two decades. The UCSC Forest Ecology Research Plot was made possible by National Science Foundation grants to G. S. Gilbert (DEB-0515520 and DEB-084259), by the Pepper-Giberson Chair Fund, the University of California, and the hard work of dozens of UCSC students. These two projects are part the Center for Tropical Forest Science, a global network of large-scale demographic tree plots. The Luquillo Experimental Forest Long-Term Ecological Research Program was supported by grants BSR-8811902, DEB 9411973, DEB

0080538, DEB 0218039, DEB 0620910 and DEB 0963447 from NSF to the Institute for Tropical Ecosystem Studies, University of Puerto Rico, and to the International Institute of Tropical Forestry USDA Forest Service, as part of the Luquillo Long-Term Ecological Research Program. Funds were contributed for the 2000 census by the Andrew Mellon foundation and by Center for Tropical Forest Science. The U.S. Forest Service (Dept. of Agriculture) and the University of Puerto Rico gave additional support. We also thank the many technicians, volunteers and interns who have contributed to data collection in the field. This research was supported by a CAREER award from the U.S. National Science Foundation (DEB-0953694) and by the Gordon and Betty Moore Foundation's Data-Driven Discovery Initiative through Grant GBMF4563, both to E. P. White.

**Table 1.** Analytical forms of the patterns predicted by the four models with interpretations. $\lambda$'s and $\lambda$''s are Lagrange multipliers for ASNE and AGSNE, respectively. $\tau$'s are parameters for SSNT_N and SSNT_M. $C$'s are normalization constants. $\gamma$ in $\Psi_{\text{ASNE}}(D)$ is defined as $\gamma = \lambda_1 + \lambda_2 \cdot D^2$, and $\gamma'$ in $\Psi_{\text{AGSNE}}(D)$ is defined as $\gamma' = \lambda_2' + \lambda_3' \cdot D^2$.

| Patterns | Species abundance distribution (SAD) | Individual size distribution (ISD) | Size-density relationship (SDR) |
|---|---|---|---|
| ASNE | $\Phi_{\text{ASNE}}(n) \approx \frac{1}{C_{ASNE1} n} e^{-(\lambda_1+\lambda_2)n}$ | $\Psi_{\text{ASNE}}(D) = \frac{D}{C_{ASNE2}} \cdot \frac{e^{-\gamma}}{(1-e^{-\gamma})^2} \cdot \left(1-(N+1)e^{-\gamma N} + Ne^{-\gamma(N+1)}\right)$ | $\bar{\varepsilon}_{\text{ASNE}}(n) = \frac{1}{n\lambda_2(e^{-\lambda_2 n} - e^{-\lambda_2 nE})} \cdot \left[e^{-\lambda_2}(\lambda_2 n + 1) - e^{-\lambda_2 nE}(\lambda_2 nE + 1)\right]$ |
| AGSNE | $\Phi_{\text{AGSNE}}(n) \approx \frac{1}{C_{AGSNE1} n} \cdot \frac{e^{-(\lambda_1'+(\lambda_2'+\lambda_3')n)}}{1-e^{-(\lambda_1'+(\lambda_2'+\lambda_3')n)}}$ | $\Psi_{\text{AGSNE}}(D) \approx \frac{D}{C_{AGSNE2}} \sum_{m=1}^{S} \frac{me^{-(\lambda_1'+\gamma')m}}{(1-e^{-\gamma' m})^2}$ | $\bar{\varepsilon}_{\text{AGSNE}}(m,n) \approx [e^{-\lambda_3' mn}(\lambda_3' mn + 1) - e^{-\lambda_3' mnE}(\lambda_3' mnE + 1)] / [\lambda_3' mn(e^{-\lambda_3' mn} - e^{-\lambda_3' mnE})]$ |
| SSNT_N | $\Phi_{\text{SSNT_N}}(n) = -\frac{1}{\ln(1-\tau_1)} \frac{{\tau_1}^n}{n}$ | $\Psi_{\text{SSNT_N}}(D) = \tau_2 e^{-\tau_2(D-1)}$ | $\bar{\varepsilon}_{\text{SSNT_N}} = \frac{2}{\tau_2^2} + \frac{2}{\tau_2} + 1$ |
| SSNT_M | $\Phi_{\text{SSNT_M}}(n) = -\frac{1}{\ln(1-\tau_1)} \frac{{\tau_1}^n}{n}$ | $\Psi_{\text{SSNT_M}}(D) = \frac{2}{3}\tau_3 e^{-\tau_3(D^{2/3}-1)} \cdot D^{-1/3}$ | $\bar{\varepsilon}_{\text{SSNT_M}} = \frac{6}{\tau_3^3} + \frac{6}{\tau_3^2} + \frac{3}{\tau_3} + 1$ |
| Interpre-tation | The probability that a randomly selected species has abundance *n*. | The probability that a randomly selected individual from the community has diameter between (*D*, *D* + Δ*D*) regardless of species identity. | The average individual metabolic rate within a species with abundance *n* (and that the species belongs to a genus with *m* species in AGSNE). Note that metabolic rate scales as $D^2$ instead of *D*. |

**Table 2.** Summary of datasets.

| Dataset | Description | Area of Individual Plots (ha) | Number of Plots | Survey Year | References |
|---|---|---|---|---|---|
| CSIRO | Tropical rainforest | 0.5 | 20 | 1985-2012* | 1 |
| Serimbu | Tropical rainforest | 1 | 1 | 1995† | 2-5 |
| La Selva | Tropical wet forest | 2.24 | 5 | 2009 | 6, 7 |
| Eno-2 | Tropical moist forest | 1 | 1 | 2000-2001 | 8 |
| BCI | Tropical moist forest | 50 | 1 | 2010 | 9-11 |
| DeWalt Bolivia forest plots | Tropical moist forest | 1 | 2 | N/A | 12 |
| Luquillo | Tropical moist forest | 16 | 1 | 1994-1996‡ | 13, 14 |
| Sherman | Tropical moist forest | 5.96 | 1 | 1999 | 15-17 |
| Cocoli | Tropical moist forest | 4 | 1 | 1998 | 15-17 |
| Western Ghats | Wet evergreen / moist / dry deciduous forests | 1 | 34 | 1996-1997 | 18 |
| UCSC FERP | Mediterranean mixed evergreen forest | 6 | 1 | 2007 | 19 |
| Shirakami | Beech forest | 1 | 2 | 2006 | 4, 5, 20 |
| Oosting | Hardwood forest | 6.55 | 1 | 1989 | 21, 22 |
| North Carolina forest plots | Mixed hardwoods / pine forest | 1.3 – 5.65 | 5 | 1990-1993§ | 23-25 |

[1]Bradford *et al.* 2014 [2]Kohyama *et al.* 2001 [3]Kohyama *et al.* 2003 [4]Lopez-Gonzalez *et al.* 2009 [5]Lopez-Gonzalez *et al.* 2011 [6]Baribault *et al.* 2011a [7]Baribault *et al.* 2011b [8]Pitman *et al.* 2005 [9]Condit 1998a [10]Hubbell *et al.* 1999 [11]Hubbell *et al.* 2005 [12]DeWalt *et al.* 1999 [13]Zimmerman *et al.* 1994 [14]Thompson *et al.* 2002 [15]Condit 1998b [16]Pyke *et al.* 2001 [17]Condit *et al.* 2004 [18]Ramesh *et al.* 2010 [19]Gilbert *et al.* 2010 [20]Nakashizuka *et al.* 2003 [21]Reed *et al.* 1993 [22]Palmer *et al.* 2007 [23]Peet & Christensen 1987 [24]McDonald *et al.* 2002 [25]Xi *et al.* 2008

[*] We chose the most recent survey in each plot before documented disturbances.
[†] One plot has a more recent survey in 1998, however it lacks species ID.
[‡] We chose Census 2 because information for multiple stems is not available in Census 3, and the unit of diameter is unclear in Census 4.
[§] We chose survey individually for each plot based on expert opinion to minimize the effect of hurricane disturbance.

**Table 3.** Joint distribution $P(n, D_1, D_2, \ldots, D_n)$ for the four models. $Z$ in AGSNE is a constant. See Table 1 for the interpretation of the other symbols and parameters, and **Appendix A** for derivations.

| Model | Predicted joint distribution |
|---|---|
| ASNE | $P_{\text{ASNE}}(n, D_1, D_2, \ldots, D_n) = \frac{1}{C_{ASNE1}n} e^{-(\lambda_1+\lambda_2)n} \prod_{i=1}^{n} \frac{2n\lambda_2 D_i e^{-\lambda_2 n D_i^2}}{e^{-\lambda_2 n} - e^{-\lambda_2 n E_{\text{METE}}}}$ |
| AGSNE | $P_{\text{AGSNE}}(n, D_1, D_2, \ldots, D_n)$ <br> $= \left(\frac{2G}{ZS}\right)^n [\Phi_{AGSNE}(n)]^{1-n} \prod_{i=1}^{n} D_i \frac{t}{(1-t)^2} [1-(S+1)t^S + St^{S+1}]$ <br> where $t(D_i) = e^{-(\lambda_1' + \lambda_2' n + \lambda_3' n D_i^2)}$ |
| SSNT_N | $P_{\text{SSNT_N}}(n, D_1, D_2, \ldots, D_n) = -\frac{1}{\log(1-\tau_1)} \frac{\tau_1^n}{n} \cdot \prod_{i=1}^{n} \tau_2 \cdot e^{-\tau_2(D_i - 1)}$ |
| SSNT_M | $P_{\text{SSNT_N}}(n, D_1, D_2, \ldots, D_n) = -\frac{1}{\log(1-\tau_1)} \frac{\tau_1^n}{n} \cdot \prod_{i=1}^{n} \frac{2}{3} \tau_3 e^{-\tau_3(D_i^{2/3} - 1)} \cdot D_i^{-1/3}$ |

## Figure Legends

**Figure 1.** Comparison of the log-likelihood (*l*) of the joint distribution $P(n, d_1, d_2, \ldots, d_n)$ for the four models in each of the 76 forest communities. *l* of AGSNE, METE_N, and METE_M are compared with that of ASNE, which has the lowest likelihood in all communities. The diagonal line is the one-to-one line. For better visualization, *l* is transformed to $-\log(-l)$, which is a monotonic transformation that does not change the position of the points with respect to the diagonal line. Note that values of *l* depend on the number of individuals within the community, and thus are not comparable across communities.

**Figure 2.** Comparison of the performance of the four models for each of the three macroecological patterns. Each point in the subplot represents the abundance of one species in a community for the SAD, the diameter of one individual in a community for the ISD, and the average metabolic rate (squared diameter) within one species in a community for the SDR. The colors represent density of the points, where warmer (redder) colors correspond to denser regions. The diagonal line represents the one-to-one line between the predicted values and the observed values.

**Figure 3.** Comparison of $R^2$ values for the three macroecological patterns predicted by the four models in each of the 76 forest communities. $R^2$ of AGSNE, METE_N, and METE_M are compared with that of ASNE. The diagonal line is the one-to-one line.

**Figure 1.**

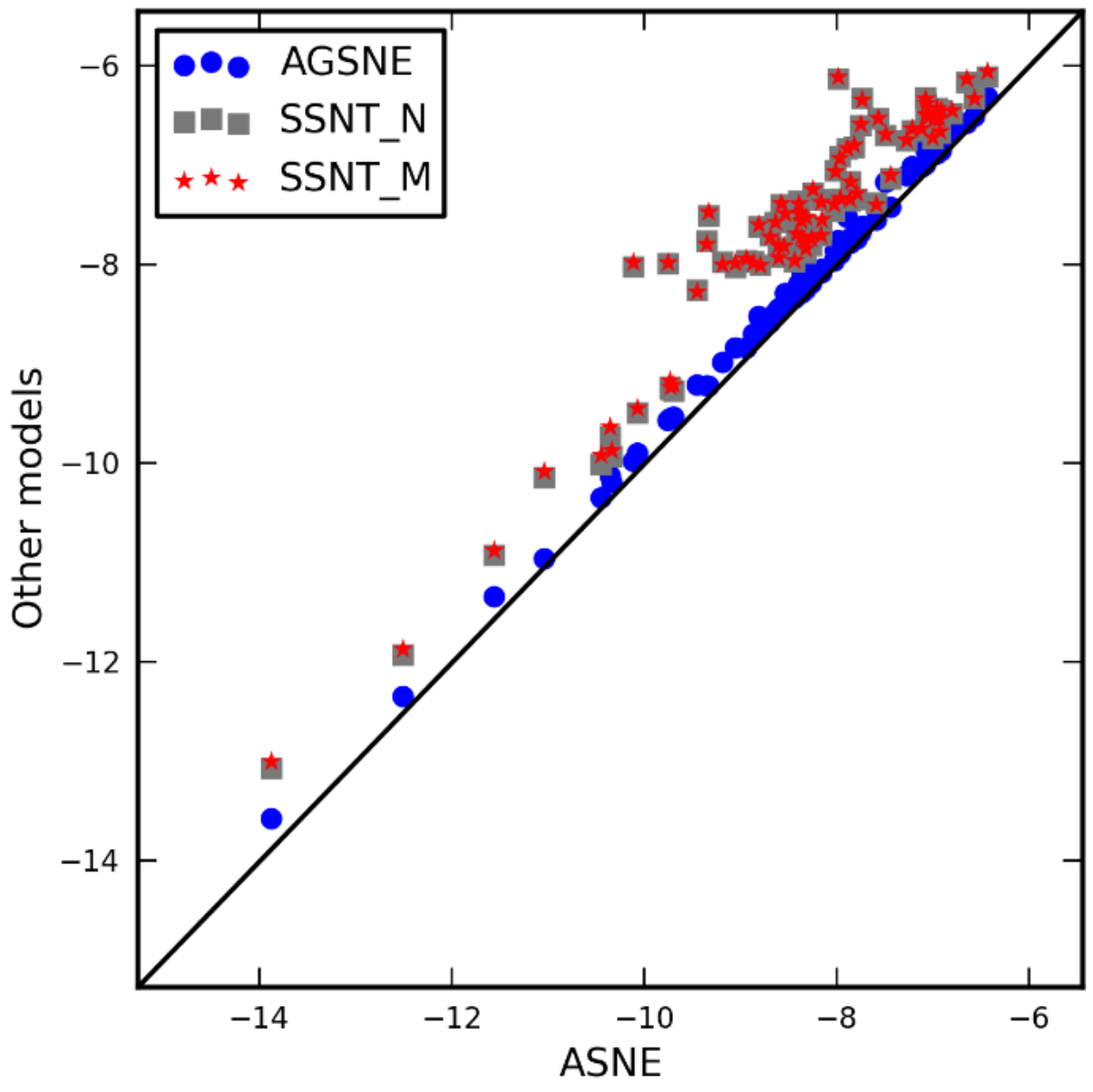

**Figure 2.**

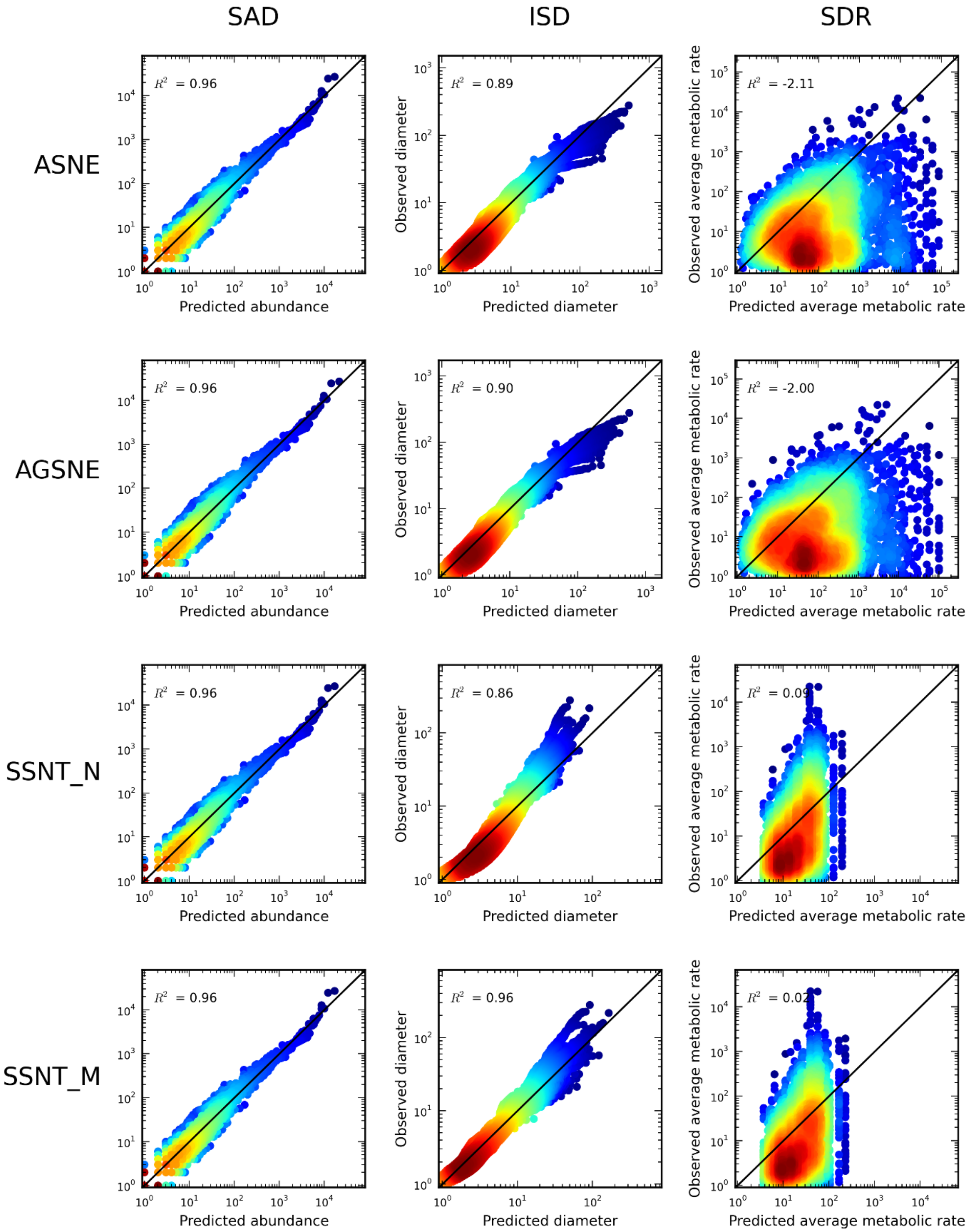

**Figure 3.**

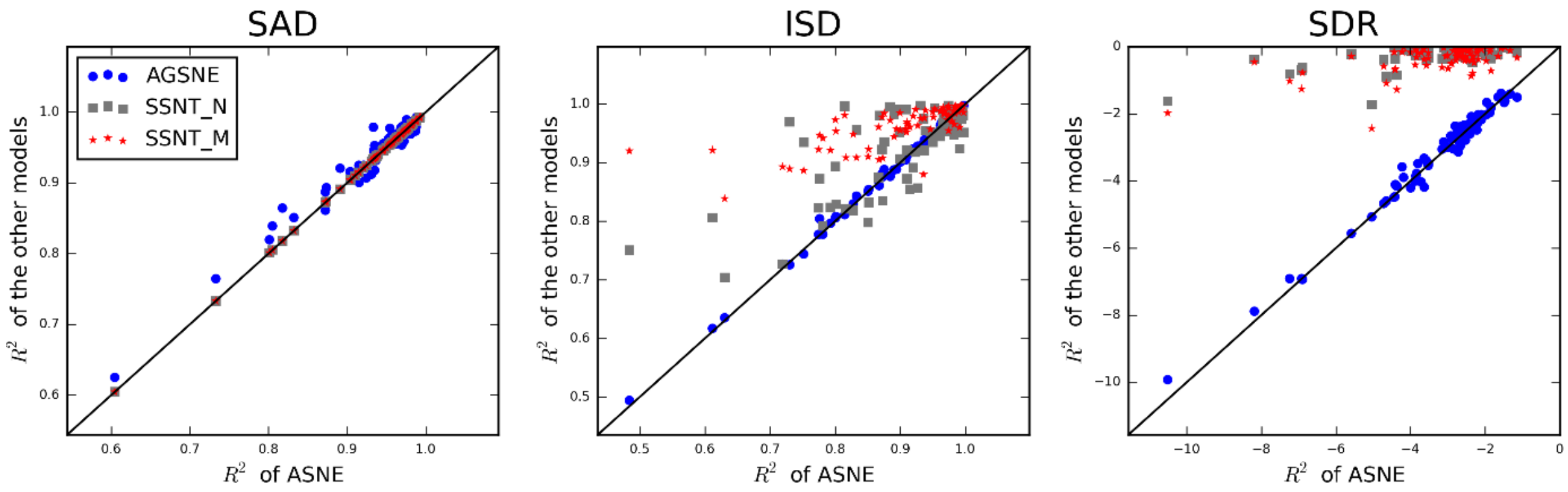

**Appendix A. Derivations**

1. ASNE

Derivations for equations in ASNE are presented in (Harte 2011) and (Xiao et al. 2015).

2. AGSNE

Derivations for AGSNE are available in Harte et al. 2015 (supplementary material). Here we only specify the equations that we used in our analysis, and give derivations for the few that are not directly presented in (Harte et al. 2015).

AGSNE is constructed with the quantity $Q(m, n, \varepsilon)$, which is the joint probability that a genus in the community has *m* species, that one species randomly chosen from these has abundance *n*, and that one randomly chosen individual from such species has metabolic rate $\varepsilon$. The constraints on the community are the average number of species, the average abundance, and the average total metabolic rate per genus, which can be expressed as expectations of $Q(m, n, \varepsilon)$:

$$\frac{S}{G} = \sum_{m,n,\varepsilon} mQ(m,n,\varepsilon) \qquad \text{(Eqn S-1 in Harte et al. 2015)}$$

$$\frac{N}{G} = \sum_{m,n,\varepsilon} mnQ(m,n,\varepsilon) \qquad \text{(Eqn S-2 in Harte et al. 2015)}$$

$$\frac{E}{G} = \sum_{m,n,\varepsilon} mn\varepsilon Q(m,n,\varepsilon) \qquad \text{(Eqn S-3 in Harte et al. 2015)}$$

where *G* is the number of genera within the community, *S* is the number of species, *N* is the total number of individuals, and *E* is the total metabolic rate across all individuals. Applying the MaxEnt algorithm yields $Q(m,n,\varepsilon) = \frac{1}{Z} e^{-\lambda_1' m} e^{-\lambda_2' mn} e^{-\lambda_3' mn\varepsilon}$ (Eqn S-4 in Harte et al. 2015), where *Z* is the normalization constant, and $\lambda_1'$, $\lambda_2'$ and $\lambda_3'$ are Lagrange multipliers given by

$$\frac{S}{G} = \frac{\sum_{m=1}^{S} e^{-\lambda_1' m} \ln(\frac{1}{1-e^{-\beta' m}})}{\sum_{m=1}^{S} \frac{e^{-\lambda_1' m}}{m} \ln(\frac{1}{1-e^{-\beta' m}})} \qquad \text{(Eqn S-17 in Harte et al. 2015)}$$

$$\frac{N}{G} = \frac{\sum_{m=1}^{S} \frac{e^{-(\lambda_1' + \beta' )m}}{1 - e^{-\beta' m}}}{\sum_{m=1}^{S} \frac{e^{-\lambda_1' m}}{m} \ln(\frac{1}{1 - e^{-\beta' m}})} \qquad \text{(Eqn S-18 in Harte et al. 2015)}$$

$$\lambda_3' = \frac{G}{E - N} \qquad \text{(Eqn S-20 in Harte et al. 2015)}$$

where $\beta' = \lambda_2' + \lambda_3'$.

AGSNE predicts the SAD as

$$\Phi_{AGSNE}(n) \approx \frac{G}{S\lambda_3' Z} \frac{1}{n} \frac{e^{-(\lambda_1' + \beta' n)}}{1 - e^{-(\lambda_1' + \beta' n)}} \approx \frac{1}{C_{AGSNE1}} \frac{1}{n} \frac{e^{-(\lambda_1' + \beta' n)}}{1 - e^{-(\lambda_1' + \beta' n)}} \qquad \text{(Eqn S-26 in Harte et al. 2015)}$$

Because approximation is used, we replaced the constant $\frac{G}{S\lambda_3' Z}$ with $1/C_{AGSNE1}$ to ensure proper normalization of $\varphi(n)$. ISD is predicted to be

$$\Psi_{AGSNE}(\varepsilon) \approx \frac{G}{NZ} \sum_{m=1}^{S} \frac{m e^{-\lambda_1' m} e^{-\gamma' m}}{(1 - e^{-\gamma' m})^2} \qquad \text{(Eqn S-40 in Harte et al. 2015)}$$

where $\gamma' = \lambda_2' + \lambda_3' \varepsilon$. Note that $\varepsilon$ is metabolic rate, and scales with diameter $D$ as $\varepsilon \propto D^2$. $\Psi_{\mathrm{AGSNE}}(\varepsilon)$ is converted into distribution of $D$ with the transformation (Casella and Berger 2001)

$$\Psi_{AGSNE}(D) = \Psi_{AGSNE}(\varepsilon) \left| \frac{d\varepsilon}{dD} \right| = \Psi_{AGSNE}(\varepsilon) \left| \frac{dD^2}{dD} \right| = 2D \cdot \Psi_{AGSNE}(\varepsilon)$$

Which yields

$$\Psi_{AGSNE}(D) \approx \frac{2G}{NZ} D \sum_{m=1}^{S} \frac{m e^{-\lambda_1' m} e^{-(\lambda_2' + \lambda_3' D^2)m}}{(1 - e^{-(\lambda_2' + \lambda_3' D^2)m})^2} \approx \frac{D}{C_{AGSNE2}} \sum_{m=1}^{S} \frac{m e^{-\lambda_1' m} e^{-(\lambda_2' + \lambda_3' D^2)m}}{(1 - e^{-(\lambda_2' + \lambda_3' D^2)m})^2} \qquad \text{(Eqn A1)}$$

where $C_{AGSNE2}$ is the normalization constant for $\Psi_{\mathrm{AGSNE}}(D)$.

The distribution of $\varepsilon$ given that the individual belongs to a species with abundance $n$ and a genus with $m$ species, is predicted as

$$\Theta_{AGSNE}(\varepsilon | m, n) = \frac{\lambda_3' m n e^{-\lambda_3' m n \varepsilon}}{e^{-\lambda_3' m n \varepsilon} - e^{-\lambda_3' m n E}} \qquad \text{(Normalized version of Eqn S-43 in Harte et al. 2015)}$$

Taking the expected value of this distribution yields the SDR

$$\bar{\varepsilon}_{AGSNE}(m,n) \approx \frac{e^{-\lambda_3' mn}(\lambda_3' mn+1)-e^{-\lambda_3' mnE}(\lambda_3' mnE+1)}{\lambda_3' mn\left(e^{-\lambda_3' mn}-e^{-\lambda_3' mnE}\right)} \quad \text{(Eqn A2)}$$

The joint distribution for abundance and body size, $P_{AGSNE}(n, D_1, D_2, \ldots, D_n)$, can be derived as:

$$P_{AGSNE}(n, D_1, D_2, \ldots D_n) = \Phi(n)P(D_1, D_2, \ldots D_n|n) = \Phi(n)\prod_{i=1}^{n} P(D_i|n)$$

(convert the joint distribution into marginal and conditional distributions)

$$= \Phi(n)\prod_{i=1}^{n} \frac{P(D_i, n)}{\Phi(n)} = [\Phi(n)]^{1-n}\prod_{i=1}^{n} 2D_i P(\varepsilon_i, n)$$

(convert distribution of diameter *D* into distribution of metabolic rate *ε*)

$$= 2^n[\Phi(n)]^{1-n}\prod_{i=1}^{n} \frac{D_i \sum_{m=1}^{S} mQ(m,n,\varepsilon)}{\sum_{m=1}^{S}\sum_{n=1}^{N}\int_{\varepsilon=1}^{E} mQ(m,n,\varepsilon)d\varepsilon}$$

(obtain P($\varepsilon_i$, *n*) by integrating out m from Q(*m*, *n*, *ε*))

$$= (\frac{2G}{ZS})^n[\Phi(n)]^{1-n}\prod_{i=1}^{n} D_i \frac{t}{(1-t)^2}[1-(S+1)t^S + St^{S+1}]$$

(Eqn A3)

where $t = e^{-(\lambda_1' + \lambda_2' n + \lambda_3' n D_i^2)}$.

2. SSNT_N and SSNT_M

Predictions of SSNT have been presented in detail in (O'Dwyer et al. 2009) for both the general case with arbitrary functional forms for the demographic parameters *b* (birth rate), *m* (mortality rate), and *g* (growth rate), where *b* is constant while *m* and *g* are allowed to vary as functions of body size, and the special case where all parameters are constant across individuals in the community (SSNT_N). We derive the form of the patterns for SSNT_N first, then obtain the predictions of SSNT_M from those of SSNT_N by transforming the unit for body size in *g*.

SSNT predicts that the size component of the theory does not affect the SAD, which still takes the same form as in Hubbell's neutral theory (Hubbell 2001):

$$\Phi_{\mathrm{SSNT_N}}(n) = \frac{\alpha}{n}{\tau_1}^n \qquad \text{(Eqn A4; Modified from Eqn 3 in O'Dwyer } et\ al. \text{ 2009)}$$

where $\alpha$ is the normalization factor. For $\Phi_{\mathrm{SSNT}}(n)$ to be properly normalized, $\alpha$ has to satisfy

$$\alpha = -\frac{1}{\ln(1-\tau_1)}$$

In addition, the total abundance in the community has to be *N*:

$$S\sum_{n=1}^{\infty} n\Phi_{SSNT_N}(n) = S\alpha\frac{\tau_1}{1-\tau_1} = N$$

Solving the above two equations simultaneously yields the value of the parameter $\tau_1$ for the SAD:

$$\frac{N}{S} = -\frac{\tau_1}{1-\tau_1}\log(1-\tau_1) \qquad \text{(Eqn A5)}$$

The mean size spectrum, or the average number of individuals per species in a given size class *D*, is given by

$$< n(D) >= \frac{\nu}{g(1-\tau_1)}e^{-\tau_2 D} \qquad \text{(Modified from Eqn 17 in O'Dwyer } et\ al. \text{ 2009)}$$

Transforming the above equation into a probability distribution (ISD) yields

$$\Psi_{\mathrm{SSNT_N}}(D) = \frac{S}{N} < n(D) >= \beta e^{-\tau_2 D} \qquad \text{(Eqn A6)}$$

where $\beta = \frac{S\nu}{Ng(1-\tau_1)}$ is the normalization factor. Similar to $\tau_1$ in the SAD, the parameter $\tau_2$ characterizing $\Psi_{\mathrm{SSNT_N}}(D)$ can be obtained by simultaneously solving the normalization equation

$$\int_{D=1}^{\infty} \Psi_{\mathrm{SSNT_N}}(D) = \frac{1}{\tau_2}\ \beta e^{-\tau_2} = 1$$

and the equation for the total diameter summed across all individuals in the community, given by

$E_{SSNT_N}$:

$$\int_{D=1}^{\infty} D \cdot \Psi_{\text{SSNT_N}}(D) = (\frac{1}{\tau_2})^2 \beta e^{-\tau_2}(\tau_2 + 1) = E_{SSNT_N}$$

Note that the lower limit of the integration comes from the fact that the observed ISD is lower-truncated at 1 after rescaling. Combining the above two equations yields

$$\tau_2 = \frac{N}{E_{SSNT_N} - N} \qquad \text{(Eqn A7)}$$

The SDR measured as average metabolic rate, or $D^2$, can then be calculated as the expected value of the $\Theta_{\text{SSNT_N}}(\varepsilon)$, the distribution of metabolic rate among individuals within a species (which in this case takes the same form as the ISD) converted to distribution of $\varepsilon = D^2$:

$$\bar{\varepsilon}_{\text{SSNT_N}} = \mathrm{E}(\Theta_{\text{SSNT_N}}(\varepsilon)) = \int_1^{\infty} \varepsilon \cdot \frac{\tau_2}{2} \cdot \frac{1}{\varepsilon^{0.5}} e^{-\tau_2(\varepsilon^{0.5}-1)} d\varepsilon$$

$$(let\ t = \varepsilon^{0.5} - 1) = \frac{\tau_2}{2} \int_0^{\infty} (t+1) e^{-\tau_2 t} d(t^2 + 2t + 1)$$

$$= \tau_2 \int_0^{\infty} (t+1)^2 e^{-\tau_2 t} dt = \frac{2}{{\tau_2}^2} + \frac{2}{\tau_2} + 1 \quad \text{(Eqn A8)}$$

The joint distribution $P_{SSNT_N}(n, D_1, D_2, \ldots, D_n)$ can be derived as

$$P_{SSNT_N}(n, D_1, D_2, \ldots D_n) = \Phi_{\text{SSNT_N}}(n) \prod_{i=1}^{n} \Psi_{\text{SSNT_N}}(D_i) = -\frac{1}{\log(1-\tau_1)} \frac{\tau_1^n}{n} \cdot \prod_{i=1}^{n} \tau_2 \cdot e^{-\tau_2(D_i - 1)}$$

(Eqn A9)

Predictions of SSNT_M can be directly obtained from those of SSNT_N, by replacing $D$ with $D^{2/3}$ and $E_{SSNT_N}$ with $E_{SSNT_M}=\sum D^{2/3}$ and transforming distributions to proper units, which yields

$$\Phi_{\text{SSNT_M}}(n) = -\frac{1}{\log(1-\tau_1)} \frac{\tau_1^n}{n} \quad \text{(Eqn A10)}$$

$$\Psi_{\text{SSNT_M}}(D) = \frac{2}{3} \tau_3 e^{-\tau_3(D^{2/3}-1)} \cdot D^{-1/3} \qquad \text{(Eqn A11)}$$

$$\bar{\varepsilon}_{\text{SSNT_M}} = \frac{6}{\tau_3^3} + \frac{6}{\tau_3^2} + \frac{3}{\tau_3} + 1 \qquad \text{(Eqn A12)}$$

$$P_{\mathrm{SSNT_M}}(n, D_1, D_2, \dots, D_n) = -\frac{1}{\log(1-\tau_1)}\frac{\tau_1^n}{n} \cdot \prod_{i=1}^{n} \frac{2}{3}\tau_3 e^{-\tau_3(D_i^{2/3}-1)} \cdot D_i^{-1/3} \quad \text{(A13)}$$

where $\tau_3 = \frac{N}{E_{SSNT_M}-N} = \frac{N}{\sum D^{2/3}-N}$.

## Appendix B. Bootstrap Analysis

In the main text we examined the performance of the models with two metrics – the log-likelihood of the joint distribution $P(n, D_1, D_2, \ldots, D_n)$, which quantifies the general performance of a model compared to another in characterizing the overall pattern of abundance and body size; and the $R^2$ value between observed values and predicted values, which quantifies the predictive power of a model for a single pattern. However, neither metric takes into account the variation resulting from finite sample sizes, which may translate into discrepancies between the observations and the predictions even when the predicted form is accurate.

Here we examine the discrepancy between random samples from a distribution and the predicted (rank) values as a measure of the intrinsic variation, which is then compared to the discrepancy between the predicted values and the observations. If the discrepancy calculated for the observations is no larger than that for the random samples, it would imply that the observations are indistinguishable from a random sample from the predicted distribution. Alternatively, if the discrepancy for the observations is significantly higher, it would imply that the observations do not fully conform to the predicted distribution.

We drew 200 random samples from the distributions predicted by each model for the SAD, the ISD, and the intraspecific individual size distribution for the SDR (see Table 1 in the main text), with the parameterization empirically obtained from the variables *G* (only used in AGSNE), *S*, *N*, and *E* ($E_{METE}$, $E_{SSNT_N}$, or $E_{SSNT_M}$). Adopting a smaller number of samples (100) did not qualitatively change our results, implying that 200 samples were sufficient for this analysis. The random samples all had the same length as the empirical data, i.e., each simulated SAD had *S* species, each simulated ISD had *N* individuals, and the SDR was obtained as the average values of the *n* individuals converted to $D^2$ for a given species.

The discrepancy between a random sample and the values predicted by the models was measured with two metrics, $R^2$ and the Kolmogorov-Smirnov (K-S) statistic. The K-S statistic takes the form

$$D_n = \sqrt{n}\,\sup|F_n(x) - F(x)| \qquad \text{(Eqn B1)}$$

where $n$ is sample size, $F_n(x)$ is the empirical cumulative distribution function, and $F(x)$ is the predicted cumulative distribution function. While $R^2$ is a measure of the explanatory power of the predictions, the K-S statistic characterizes the overall difference in shape between the observed and the predicted distributions.

We computed the $R^2$ for all three patterns, and the K-S statistic for the SAD and the ISD, because the SDR is not a probability distribution and thus the K-S statistic does not apply. We compared the statistics obtained for empirical observations to those obtained for random samples of the predicted distributions by calculating the proportion (quantile) of random samples that have **equal or higher discrepancy** (i.e., lower values of $R^2$ or larger K-S statistic) than the observations.

As Figs B1 – B4 show, the SAD predicted by all models provides a satisfactory characterization of the empirical distribution of abundance among species in the majority of communities (i.e., a non-negligible proportion of random samples show equal or higher discrepancy compared to the observed values). The empirical patterns of the ISD differ from the predictions of all models, with SSNT_M doing marginally better than other models. However, SSNT_N and SSNT_M have significantly better fit for the SDR, where the pattern in most communities is indistinguishable from random samples. This implies that SSNT's prediction of no correlation between body size and species abundance may be more or less on target, despite the fact that the empirical ISD does not conform to the predicted exponential distribution (Table 1 in in the main text).

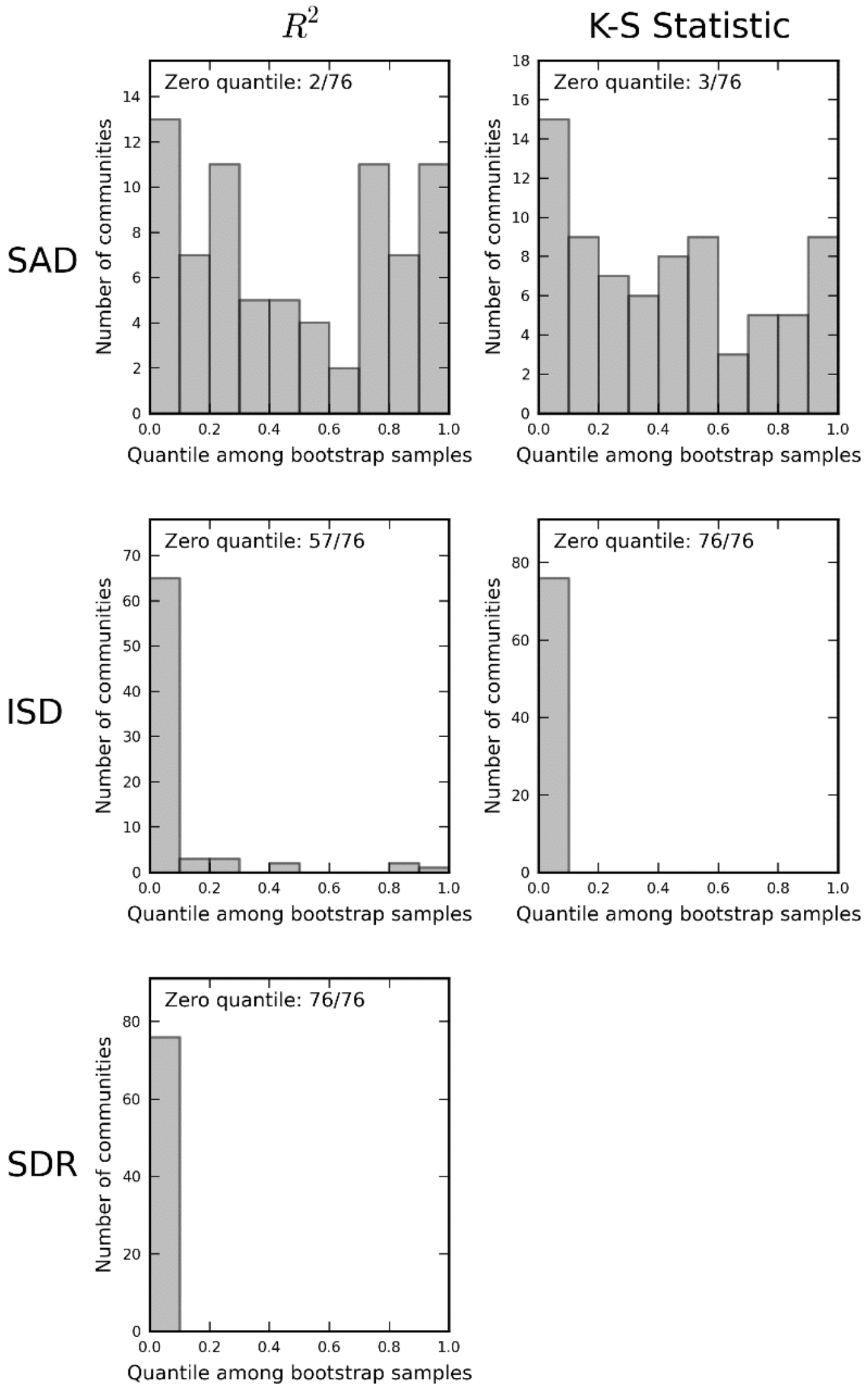

**Figure B1.** Results of the bootstrap analysis for ASNE. The histogram in each panel shows the frequency distribution of bootstrap results among the 76 communities for a given pattern, where each value represents the proportion of random samples (among 200) that are closer to the predicted values (having lower $R^2$ or larger K-S statistic) compared to the observations. The text gives the number of communities (out of 76) with zero quantiles, which indicates that the predictions are nowhere close in characterizing the empirical data for the given pattern.

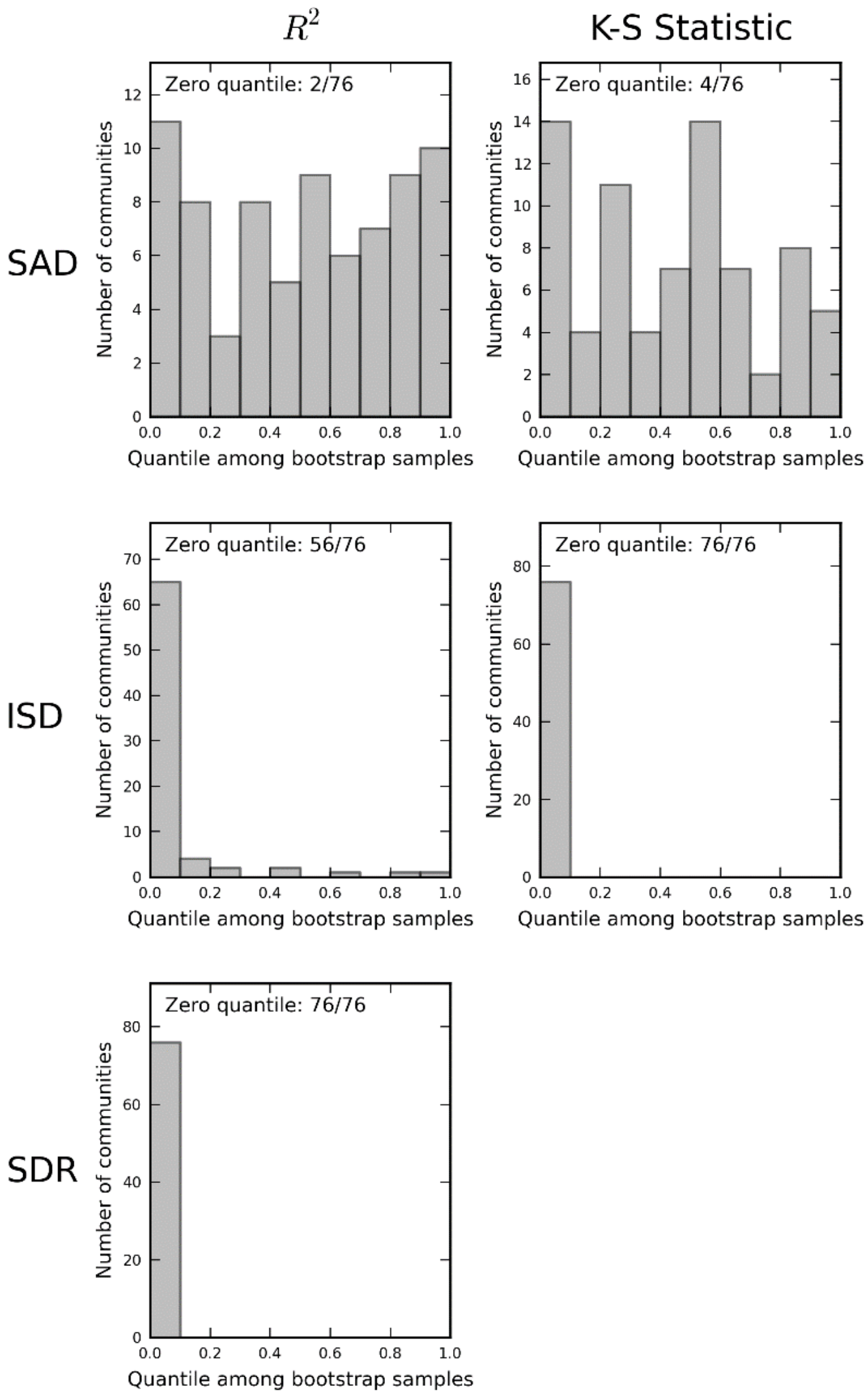

**Figure B2.** Results of the bootstrap analysis for AGSNE.

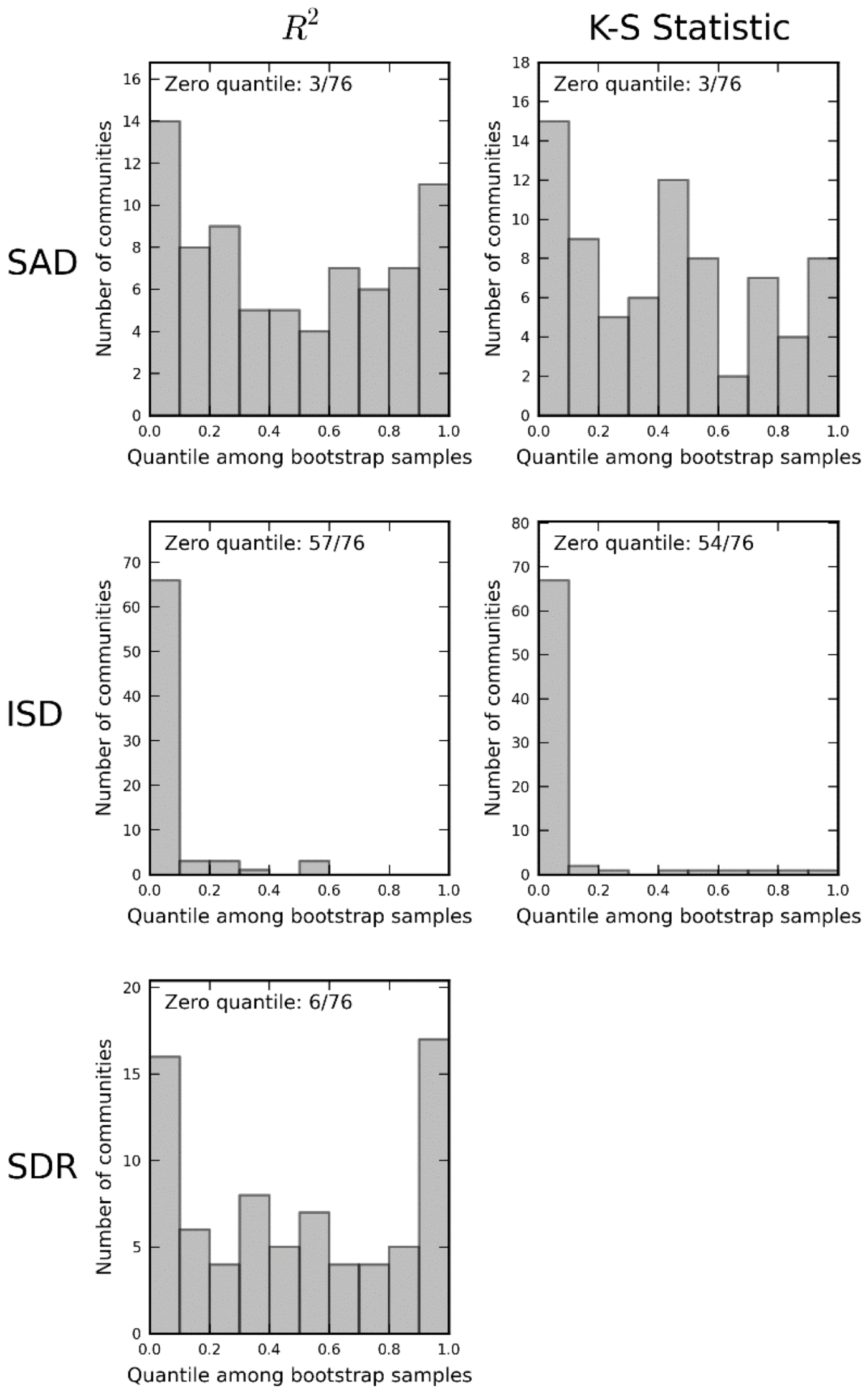

**Figure B3.** Results of the bootstrap analysis for SSNT_N.

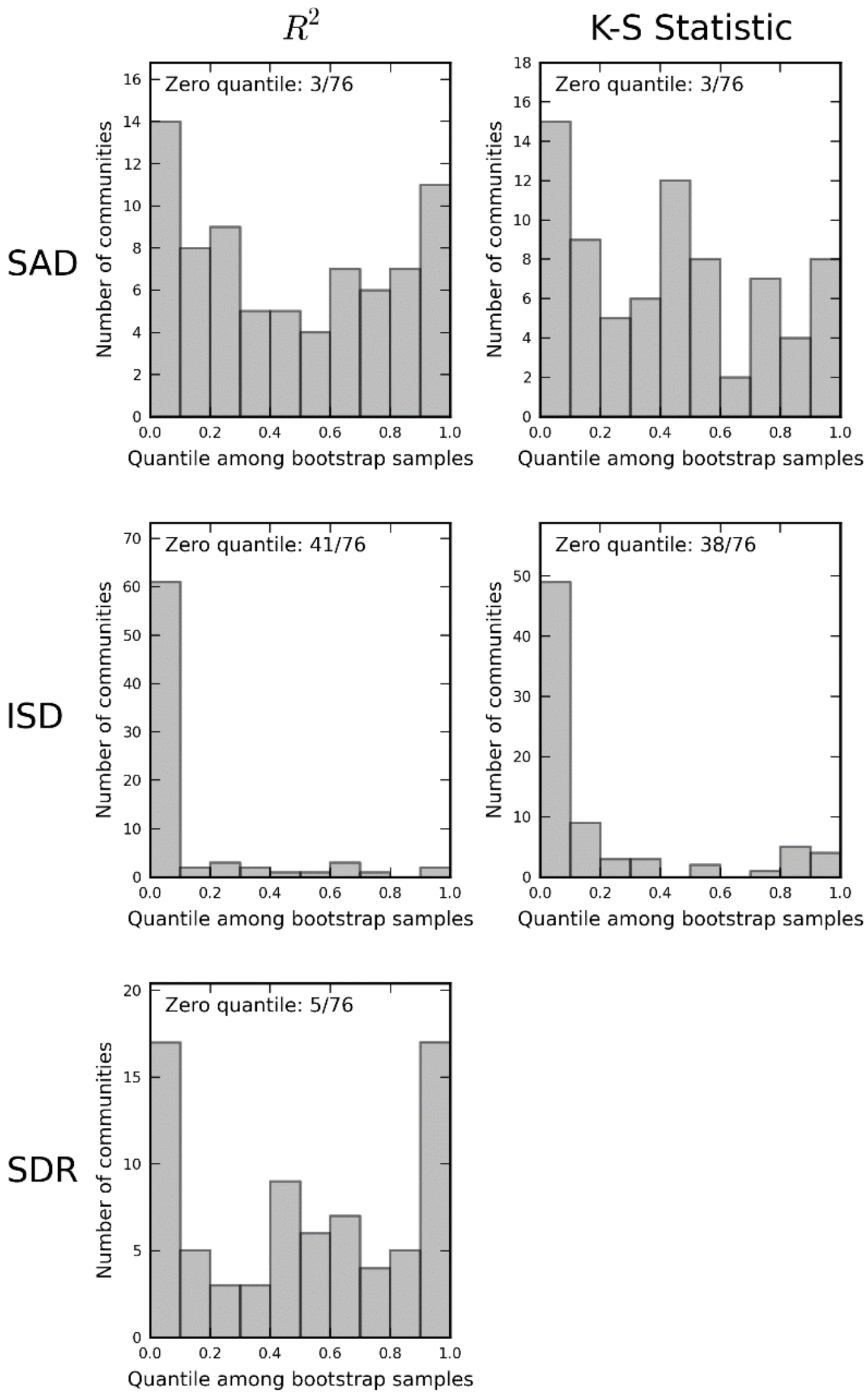

**Figure B4.** Results of the bootstrap analysis for SSNT_N.